\documentclass{article}
\usepackage{hyperref}
\usepackage{amssymb}
\usepackage{graphicx}
\usepackage{pdflscape}
\usepackage{caption} 

\newcommand{\be}{\begin{equation}}
\newcommand{\ee}{\end{equation}}
\newcommand{\ba}{\begin{eqnarray}}
\newcommand{\ea}{\end{eqnarray}}
\title{A thrust to trust minimum thrust}
\author{Matteo Cacciari\thanks{cacciari@lpthe.jussieu.fr}\\
\small\it
Sorbonne Universit\'e, CNRS, Laboratoire de Physique Th\'eorique et Hautes \'Energies, LPTHE,
F-75005 Paris, France\\
\small\it
Universit\'e Paris Cit\'e, LPTHE, F-75006 Paris, France}

\date{}

\begin{document}

\maketitle

\begin{abstract}
    We determine the minimum value of thrust for a number of $N$-particle configurations. For $N=5$ in three dimensions an exact result is found for the first time. For larger $N$ we obtain numerical results through optimisation. When a definite value cannot be reliably identified, the results are analysed in the context of Extreme Value Theory, using a Maximum Likelihood Estimate and a Bayesian analysis. Results are given for three spatial dimensions, two dimensions, and selected cases in $d$ dimensions.
\end{abstract}

\section{Introduction}

In high energy physics events produced at colliders are often studied in terms of {\sl event shapes}, functions of the three-momenta of the particles produced in a collision. These event shapes can be used to extract information about the structure of an event and are often calculable from theory, hence permitting theory-experiment comparisons. One such shape is the {\sl thrust}~\cite{Brandt:1964sa, Farhi:1977sg}.

The thrust of a set of 3-momenta $\{p_i\}$ can be defined as
\be
\label{eq:thrust}
T = \max_n \frac{\sum_i|p_i\cdot n|}{\sum_i|p_i|} \qquad \mathrm{with}~~ |n|=1
\ee
We will consider a set-up where momentum conservation gives $\sum_i p_i = 0$.
The thrust is maximized ($T=1$) when all the momenta are parallel or anti-parallel, and is minimized ($T=1/2$) when an infinite number of momenta is distributed isotropically.

The problem that we want to address here is to find $T^{min}_N$, the minimum value of thrust for a configuration consisting of a finite number $N$ of momenta $\{p_i\}$, with $i=1,\ldots,N$. We will express our results in terms of the equivalent variable $\tau_N = 1-T_N$, looking for its maximum $\tau^{max}_N$:
\be
\label{eq:taumax}
\tau^{max}_N = 1-T^{min}_N = 1 - \min_{\{p_i\}} T_N = 1-  \min_{\{p_i\}} \max_n \frac{\sum_i|p_i\cdot n|}{\sum_i|p_i|}
\ee

Solutions for $\tau^{max}_N$ are know analytically for $N=3$, where $\tau^{max}_{N=3} = 1/3 \simeq 0.333333$ for the configuration with three momenta corresponding to the vertices of an equilateral triangle inscribed in a circle, and for $N=4$, where $\tau^{max}_{N=4} = 1-1/\sqrt{3}\simeq 0.4226497$ for the configuration with four momenta corresponding to the vertices of a regular tetrahedron.

Beyond $N=4$, to our knowledge no analytical solutions are known. Ref.~\cite{Aglietti:2025jdj} has recently attempted to obtain numerically the values for $\tau^{max}_N$ for $N$ up to 14, confirming the known values for $N=3,4$, and improving a previous estimate for $\tau^{max}_{N=5}$ given in ref.~\cite{Monni:2011gb}, pushing up the predicted value from 0.4275 to 0.454. Ref.~\cite{Aglietti:2025jdj} has employed stochastic optimisation algorithms, like a genetic algorithm and a particle swarm algorithm. As its authors have carefully pointed out, such algorithms are not guaranteed to find the absolute global maximum  for $\tau_N$, because of their stochastic nature. The values that they quote for $\tau^{max}_N$ should therefore be viewed, potentially, as lower bounds only.

The purpose of this note is to present an independent determination of the values of $\tau^{max}_N$ and cross-check the results of ref.~\cite{Aglietti:2025jdj}. In the process, we are also able to determine for the first time an exact result for $N=5$ particles, given in Appendix~\ref{app:n5conf}.

\section{The numerical set-up}

Observing the known results for $N=3,4$, one may naively expect the configurations leading to $\tau^{max}_N$ to be regular polyhedrons, possibly with all the vertices lying on the unit sphere. By extrapolating what happens for $N=4$, where a tetrahedral configuration gives $\tau^{max}_{N=4}$, one may further expect the maximal configurations for $N=6, 8, 12, 20$ to be given by the platonic solids. Such expectations turn out to be misguided. For instance, for $N=6$ a regular octahedron would give the analytically calculable value $\tau^{octahedron}_{N=6} = 1-1/\sqrt{3}\simeq 0.4226497$, which is smaller than the $\tau^{max}_{N=6} = 0.463$ reported in ref.~\cite{Aglietti:2025jdj}. This suggests that it is very unlikely that one can identify simple configurations leading to $\tau^{max}_N$, possibly allowing these values to be calculated analytically. Rather, a numerical approach like the one pursued by ref.~\cite{Aglietti:2025jdj} seems to be needed.\footnote{A partial (since the determination has been made easier by first knowing the result numerically) exception is the $N=5$ case, for which an exact result is given in Appendix~\ref{app:n5conf}.}

The quantity $\sum_i|p_i\cdot n|$ that enters the calculation of $\tau^{max}_N$ in eq.~(\ref{eq:taumax}) is not a convex function. This prevents the application of minimax theorems,
which in other circumstances may have permitted the exchange of the max and of the min and the exploration of other avenues for the calculation, and seems to suggest that there is no obvious alternative to calculating numerically and successively the max and the min, in the order they are written in eq.~(\ref{eq:taumax}).

\subsection{The calculation of $T_N$}

The calculation of the thrust function is well known in the literature. While in principle it can be solved as an optimisation problem, as written in eq.~(\ref{eq:thrust}), this is slow and, more importantly in this context, usually not sufficiently accurate. Fortunately, algorithmic alternatives exist. In the case of $N=3$, explicit results are known~\cite{Banfi:2000si}: the thrust axis coincides with the momentum with the largest module. For larger $N$, an exact algorithm is given in Appendix A of \cite{Weinzierl:2009ms}. We describe it here for completeness, and give more details in our own Appendix~\ref{app:thrust}.

Given $N$ momenta $\{p_i\}$, with $\sum_i p_i=0$,  construct a set $S$ of all patterns of $N$ signs $s_i \in \{-1,1\}$ with $i=1,...,N$. The desired maximum in eq.~(\ref{eq:thrust}) is given by
\be
\label{eq:thrustalg}
\max_n \sum_i|p_i\cdot n| 
=  \max_{S} |\sum_i s_i p_i|
\ee
The thrust axis is given by $n_T = \sum_j s_j^T p_j / |\sum_j s_j^T p_j|$, where $\{s_j^T\} \in S$ is the pattern that maximises the function above. 
The search for a maximum over a continuous direction of the axis $n$ has therefore been replaced by a search over a finite set $S$ of sign patterns.

The following implementation of this algorithm in Python with \texttt{NumPy} \cite{numpy} is particularly efficient:\footnote{This is the case as long as the number $N$ of momenta is not so large that the \texttt{signs} array, whose size grows like $N (2^{N-1} -1)$, overflows the available RAM. 
Beyond this number, a `chunked' variant of the algorithm can be implemented. If $N\le 32$ \texttt{dtype=np.uint64} can be replaced with \texttt{dtype=np.uint32}. Note also that some of the array shapes specified in the comments to the code are specific to the case of 3-vectors, but the algorithm is general.}
\begin{verbatim}
    import numpy as np

    # P is an array of N 3-vectors, of shape (N,3)

    # Number of patterns = 2^(N-1)
    # Only half the patterns, since thrust is invariant 
    # under n -> -n
    n_patterns = 1 << (N-1)

    # Generate n_patterns-1 for the last N-1 signs. 
    # This excludes the combination with only +1: 
    # with the +1 prepended below, it would lead to a 
    # zero-norm because of momentum conservation. 
    # Shape (2^(N-1)-1, N-1)
    signs = 1 - 2 * (((np.arange(1, n_patterns,
                       dtype=np.uint64)[:, None] 
                       >> np.arange(N-1), 
                       dtype=np.uint64) & 1).astype(np.int8))

    # Prepend a column of +1 to fix the first sign. 
    # Shape (2^(N-1)-1, N)
    signs = np.hstack([np.ones((n_patterns-1, 1), 
                       dtype=np.int8), signs])

    # Compute all candidate vectors. Shape (2^(N-1)-1, 3)
    V = signs @ P

    # Norms of each candidate vector. Shape (2^(N-1)-1)
    norms = np.linalg.norm(V, axis=1)

    # Maximum norm
    max_norm = norms.max()
\end{verbatim}

Despite its efficiency, this algorithm still scales like $2^{N-1}-1$, and eventually becomes very slow at large $N$. Empirically, for $N \gtrsim 15$ an iterative algorithm is faster:\footnote{This cross-over point has been observed on a Mac with an M2 processor and 16GB of RAM, and working with 3-vectors.}
\begin{enumerate}
    \item 
    Generate a random axis $n^{(0)}$ as a unit vector.
    \item
    As step $k$ of an iteration, calculate the set of signs
    \be
    s_i^{(k)} = \frac{p_i\cdot n^{(k)}}{|p_i\cdot n^{(k)}|}  \qquad \forall i
    \ee
    \item
    Calculate a new axis
    \be
    n^{(k+1)} = \frac{\sum_i s_i^{(k)} p_i}{|\sum_i s_i^{(k)} p_i|}
    \ee
    and the thrust value
    \be
    T_N^{(k+1)} =  \frac{\sum_i|p_i\cdot n^{(k+1)}|}{\sum_i|p_i|}
    \ee
    \item
    Repeat this procedure from step 2 until convergence below a given tolerance, either of the axis, or of the thrust value, or of both.
\end{enumerate}
Since the function to be minimised is non-convex, to mitigate the risk of converging to a local minimum one can restart the process with a different initial random axis a number of times, and select the best result. In practice, we have used a maximum of 1000 iterations and 64 restarts.

\subsection{The calculation of $\tau^{max}_N$}

The external minimisation in eq.~(\ref{eq:taumax}) must be performed over the components of the $N$ momenta. In three dimensions, choosing arbitrarily a direction and a scale eliminates three degrees of freedom, and momentum conservation eliminates three more. We are therefore left with performing an optimisation over $3(N-2)$ parameters. Because of the highly non-convex character of the function to be minimised, global optimisation algorithms are needed. Stochastic evolutionary methods have been found empirically to work best for our problem. 

We have used the implementation of the DE (Differential Evolution) algorithm \cite{de} from the \texttt{SciPy} package \cite{scipy} in Python with the following parameters:
\begin{verbatim}
    tolerance=1e-14      (or tolerance=1e-8, depending on N)
    popsize=100          (or popsize=200)
    strategy='best1bin'
    mutation=(0.5, 1.0)
    recombination=0.95
\end{verbatim}

We have further used the CMA-ES (Covariance Matrix Adaptation Evolution strategy) algorithm \cite{cmaes} from the \texttt{Pymoo} package \cite{pymoo} (which uses the implementation available from \texttt{PyPi} \cite{pypi}), with the following parameters:
\begin{verbatim}
    tolerance=1e-8      (or tolerance=1e-6)
    pop_size=3*(N-2)    (or pop_size=10*(N-2))
    sigma=1             (or sigma=2)
    restarts=6
\end{verbatim}

In both cases, we have polished the best fit result returned by these algorithms by running the minimiser \texttt{TNC} from \texttt{SciPy}. 

We have also tested Dual Annealing and Basin Hopping from \texttt{SciPy}, as well as Genetic Algorithm and Improved Stochastic Ranking Evolutionary Strategy from \texttt{Pymoo}. We make no claim that the algorithms and the parameters that we eventually used represent the best possible choices. They were found to work sufficiently well after multiple manual tests.

\section{Statistical analysis and results}

Because of their stochastic nature, the DE and the CMA-ES algorithms are not guaranteed to find the true global minimum in a finite amount of computer time. We have therefore adopted, on a case-by-case basis, specific strategies and performed statistical analyses in order to evaluate our results.

For $N=3, 4$ and $5$ at least one of the stochastic minimisers always converge to a unique value at the tolerance level that has been set. These results coincide to all decimal figures with the known exact values $\tau^{max}_3 = 1/3$ and $\tau^{max}_4 = 1 - 1/\sqrt{3}$, and with the new $\tau^{max}_5$ given in Appendix \ref{app:n5conf}.

For larger values of $N$ the convergence becomes increasingly more difficult, and one observes a distribution of values for $\tau^{max}_N$.
In this latter case, the critical question becomes whether the largest observed value represents the true maximum, or if further runs of the stochastic code might reveal even larger values. We have tackled this question with the following approach.

After having collected the results from multiple runs (at least several tens, or a few hundred) using the same minimiser and the same parameters, we first check if, at decreasing rounding levels (lower than the tolerance that has been set for the runs) the largest value is observed multiple times. If this is the case, it is likely (though formally not certain) that it is the putative $\tau^{max}_N$ with the best possible accuracy.

The minimisation problem becomes quickly harder as $N$ increases, and the observation of many occurencies of the same value at high accuracy become rarer: observing the same maximum a few times only may not guarantee with high confidence that it represents the absolute maximum. In such cases we have therefore attempted a statistical analysis.

Our analysis rests on Extreme Value Theory and the Fisher-Tippett-Gnedenko theorem,
which states that the maximum of a sample of independent and identically distributed random variables\footnote{These conditions may not be fully satistfied in this setting. The existence of a handful of deep local minima, which the minimiser may be more likely to reach, could skew and shape the distribution. For the sake of simplicity, we will ignore this complication.} converges to three possible distribution families: the Gumbel, the Fr\'echet, or the Weibull distribution. They are collectively known as the generalised extreme value distribution. The Weibull distribution has a finite endpoint $x_{max}$ and is therefore the one that we will use. Its probability density function can be written as
\ba
\mathrm{PDF}(x; \mu,\sigma,\xi) &=& \frac{1}{\sigma} t(x)^{\xi+1} \exp(-t(x)) \qquad \nonumber\\
&&\mathrm{with}~ t(x) = \left[1 + \xi\left(\frac{x-\mu}{\sigma}\right) \right]^{-1/\xi} \quad \mathrm{and}~\xi < 0
\ea
and has support $x\in [-\infty, x_{max} = \mu - \frac{\sigma}{\xi}]$. The three parameters $\mu$, $\sigma$ and $\xi$ are often known as location, scale and shape respectively.

\begin{table}[t]
\begin{center}
    \small
    \begin{tabular}{l l c c l l c c}
        \hline
        $N$ & Observed $\tau^{max}_N$ 
            & \begin{tabular}[c]{@{}c@{}}Multiple\\observations\end{tabular} 
            & Minimiser
            & \begin{tabular}[c]{@{}c@{}}MLE fit\\for endpoint\end{tabular}
            & \begin{tabular}[c]{@{}c@{}}95\% upper\\credible bound\end{tabular}
            & \begin{tabular}[c]{@{}c@{}}68\% smallest\\credible interval\end{tabular}
            & Ref.~\cite{Aglietti:2025jdj} \\
        \hline
        3           & 0.33333333333333 & --      & DE       & --            &   --     & --                   & --    \\
        4           & 0.42264973081037 & --      & DE       & --            &   --     & --                   & --    \\
        5           & 0.45399486580028 & 100/100 & DE       & --            &   --     & --                   & 0.454 \\
        6           & 0.46365344258188 & 42/100  & DE       & --            &   --     & --                   & 0.463 \\
        7           & 0.47402556454109 & 10/50   & DE       & --            &   --     & --                   & 0.472 \\ 
        8           & 0.4794393476107  & 56/500  & DE       & --            &   --     & --                   & 0.475 \\ 
        9$^{(*)}$   & 0.4834291842117  & 12/500  & DE       & 0.48361(7)    & 0.483678 & [0.483530, 0.483627] & 0.479 \\ 
        9$^{(*)}$   & 0.483429184      & 11/500  & CMA-ES   & --            &   --     & --                   & 0.479 \\
        10$^{(*)}$  & 0.485274995845   & 5/500   & DE       & 0.48536(5)    & 0.485417 & [0.485307, 0.485372] & 0.481 \\
        10$^{(*)}$  & 0.485274996      & 3/500   & CMA-ES   & 0.485286(7)   & 0.485304 & [0.485279, 0.485292] & 0.481 \\
        11$^{(*)}$  & 0.48761809234    & 2/100   & DE       & 0.48768(12)   & 0.488127 & [0.487684, 0.487929] & 0.483 \\
        11$^{(*)}$  & 0.48761809       & 6/1000   & CMA-ES   & 0.48771(2)    & 0.487754 & [0.487672, 0.487726]& 0.483 \\
        12$^{(*)}$  & 0.4886904        & NO/100  & DE       & 0.48875(9)    & 0.488914 & [0.488705, 0.488811] & 0.484 \\
        12$^{(*)}$  & 0.4886904        & NO/500  & CMA-ES   & 0.48872(2)    & 0.488768 & [0.488699, 0.488736] & 0.484 \\
        13          & 0.4898755        & NO/500  & CMA-ES   & 0.48984(8)    & 0.489926 & [0.489878, 0.489901] & 0.485 \\
        14          & 0.4908994        & NO/500  & CMA-ES   & 0.49088(5)    & 0.490958 & [0.490904, 0.490932] & 0.486 \\ 
        15          & 0.4925471        & NO/500  & CMA-ES   & 0.49260(24)   & 0.492825 & [0.492604, 0.492730] & --    \\
        20          & 0.4940428        & NO/100  & CMA-ES   & 0.49410(10)   & 0.494321 & [0.494046, 0.494174] & --    \\
        \hline
    \end{tabular}
    \caption{\label{table:taumax}$\tau^{max}_N$ from numerical optimisations for $N$-particle configurations in three dimensions.
    The ``MLE fit'' values for the endpoint are obtained with a Maximum Likelihood Estimate, with the number in brackets representing the 1-sigma uncertainty on the last decimal figure(s). The 95\% upper credible bounds and the 68\% smallest credible intervals (SCI) are obtained with a Bayesian analysis. The dashes denote cases where the statistical analysis does not point to anything different from the already observed $\tau^{max}_N$. Previous results from Ref.~\cite{Aglietti:2025jdj} are listed in the last column.\\ $^{(*)}$ See also footnote \#\ref{foot:n9-10-11} concerning the $N=9$, 10, 11 and 12 cases.}
\end{center}
\end{table}

\begin{figure}[t]
\includegraphics[width=0.325\textwidth]{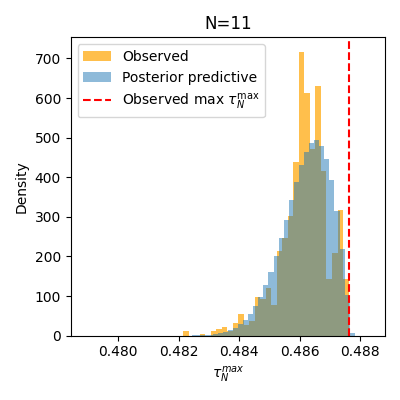}
\includegraphics[width=0.325\textwidth]{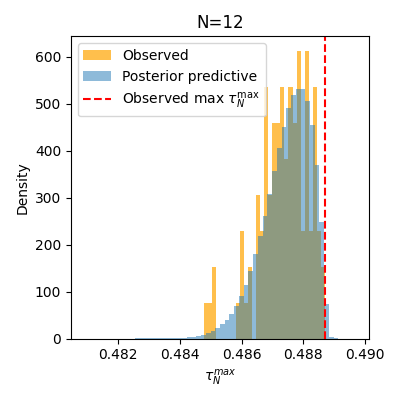}
\includegraphics[width=0.325\textwidth]{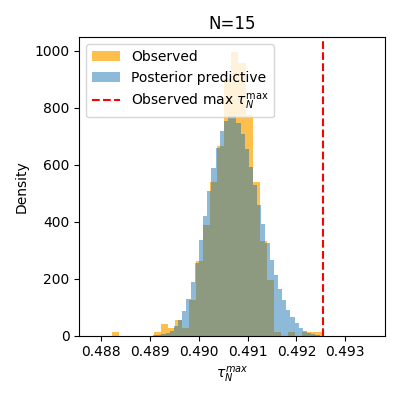}
\includegraphics[width=0.325\textwidth]{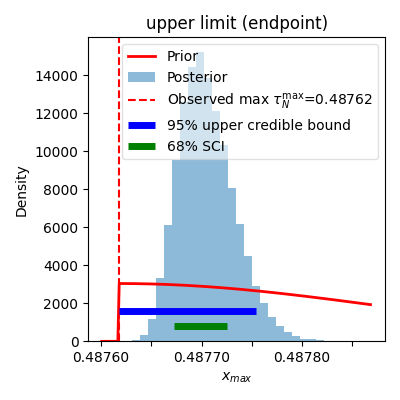}
\includegraphics[width=0.325\textwidth]{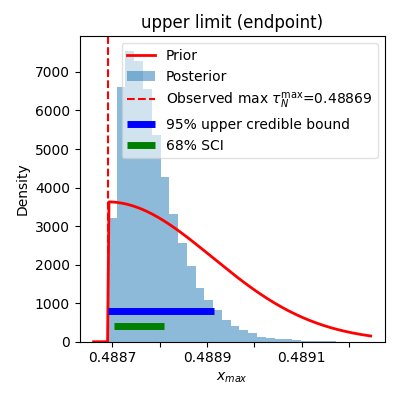}
\includegraphics[width=0.325\textwidth]{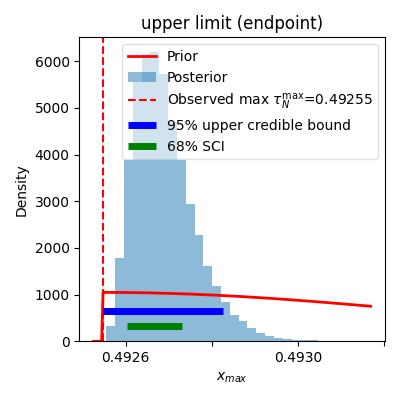}
\caption{\label{fig:bayes}Observed and posterior predictive distribution for the maxima (top row), and prior and posterior distribution for the endpoint of the inferred Weibull distribution (bottom row). We have used 1000 runs of the CMA-ES minimiser for $N=11$, 100 runs of DE for $N=12$, and 500 runs of CMA-ES for $N=15$. 
}
\end{figure}

One can fit the Weibull function to an observed distribution, trying to extract the values for the endpoint. We show results from this approach in Table~\ref{table:taumax}, with fits performed with a Maximum Likelihood Estimate (MLE) and the uncertainty obtained with the bootstrap method.

A more refined approach is to use the Weibull function as the likelihood for a Bayesian analysis. We have implemented such an analysis in \texttt{PyMC}~\cite{pymc} using the following priors for the parameters $\mu$, $\sigma$ and $\xi$:
\ba
&&\mu = {\cal N}(\bar \tau, \bar \tau \times 10^{-4}) \\
&&\sigma = {\cal N}(\sigma_\tau, \sigma_\tau \times 10^{-1}) \\
&&\label{eq:prior-xmax}\xi  = - \frac{\sigma}{x_{max} - \mu} \qquad \mathrm{with}~x_{max} =  {\cal H}(\tau_{max}, \frac{(\bar\tau - \tau_{max})^2}{\sigma_\tau}\times 10^{-1})
\ea
where ${\cal N}(\mu,\sigma)$ is a Normal distribution, ${\cal H}(\mu,\sigma)$ a half-Normal (only values larger than the mean of the corresponding Normal distribution are used\footnote{This means, of course, that the parameter $\mu$ is not the mean of the half-Normal, but  its lower limit.}), $\bar \tau$ is the average of the observed distribution of maxima, $\sigma_\tau$ their standard deviation, $\tau_{max}$ the largest observed maximum.
One can observe that we have reparametrised the problem, setting a prior on the value of the endpoint $x_{max}$ of the likelihood rather than on its shape parameter $\xi$, which becomes instead a deterministic derived quantity. While this prior is more prescriptive, it appears entirely justified, since a $\tau_N$ value as large as $\tau_{max}$ has certainly been observed and sets a fully known lower value for the endpoint.\footnote{The non-intuitive choice for the width of the half-Normal distribution in the prior for the upper limit in eq.~(\ref{eq:prior-xmax}) has been made so that it roughly corresponds to the choice of a width of $10^{-1}$ in a Normal distribution prior directly applied to the $\xi$ parameter instead. It has been checked that the posteriors obtained from using a prior on $\xi$ or a prior on the upper limit $x_{max}$ are largely equivalent, but the Markov chain Montecarlo usually converges more quickly when the upper limit is used. It has also been checked that the results of the Bayesian analysis are stable with respect to reasonable variations of all the priors.}

From this set-up, and from the distributions obtained running the stochastic minimisers, we evaluate the posteriors using \texttt{PyMC}, and the credible intervals for the true $\tau_N^{max}$.

Figure \ref{fig:bayes} shows, in the top row, the distribution of the observed maxima for a selected choice of number of particles, as well as the posterior predictive obtained after inference. The bottom row shows instead the prior and the posterior on the position of the endpoint beyond the observed maximum. The ``95\% upper credible bound'' and the ``68\% smallest credible interval'', also shown in these figures, are calculated from these posteriors on the endpoint value.

Table \ref{table:taumax} collects the maxima that we have determined, either as direct observations, or as fits or credible bounds or intervals obtained through the statistical analyses from the observed distributions.\footnote{\label{foot:n9-10-11}It is worth noting the intrinsic ambiguity of the $N=9,10,11$ and $12$ cases. For $N=9,10,11$ the maximum value has been observed multiple times, but the MLE and the Bayesian statistical analyses tend to point to a larger true maximum. In the $N=12$ case the maximum has not been observed multiple times in the individual DE and CMA-ES runs reported in the table, but the two minimisers have found the same maximum. Ultimately, a more detailed numerical analysis would likely be required to definitively determine whether the true maximum has already been observed in these cases.}

\section{Conclusions}

We have evaluated numerically the minimum kinematically allowed value of thrust $T_N$ (or, equivalently, the maximum value of $\tau_N = 1-T_N$) for $N$-particle configurations up to $N=20$. Numerical results are given in Table~\ref{table:taumax}.
For $N=3,4$ we confirm the analytically known values with 14 decimal figures. 
For $N=5$ we obtain for the first time an exact result (see Appendix~\ref{app:n5conf}), and we confirm the numerical value found in ref.~\cite{Aglietti:2025jdj}. 
For $N$ between 6 and 8 we determine numerically what are very likely the true values of $\tau_N^{max}$.
For $N$ up to 12,  we determine numerically values for $\tau_N^{max}$, but with low and progressively decreasing levels of confidence. 
For $N$ larger than 12, and up to $N=20$, we estimate credible intervals for the true maxima.
Our numerical results are slightly larger (from about 0.14\% at $N=6$ up to almost 1\% at $N=14$) than those reported in \cite{Aglietti:2025jdj}, superseding them as best estimates of $\tau_N^{max}$.
However, we do not expect such small differences to have major consequences at the phenomenological level.
Exact results in two dimensions and for selected cases in $d$ dimensions are also given in the Appendices.
An intriguing outcome of this work is the observation that, in many cases, the configuration of minimum thrust is less isotropic than one may naively expect.

\section*{Acknowledgements}
I thank Giancarlo Ferrera, Jiahao Miao, Daniele Atzori for useful conversations, exchange of results, reading of the manuscript. I am grateful to Stefan Weinzierl for exchanges on the exact thrust algorithm and the small extension presented in this paper.
Conversations with Microsoft Copilot (\url{https://copilot.microsoft.com/}), mostly with the GPT-5 model, are acknowledged. No original finding reported here was provided directly by Copilot, but references to the literature and computer code - thoroughly tested, cross-checked and verified - have been found to be helpful, with the main added value having been code suggestions and optimisations. Should any error remain, responsibility rests entirely with the author.
%

\appendix

\section{The exact thrust algorithm}
\label{app:thrust}
We reproduce here the derivation of the exact algorithm for determining the thrust value and axis given in \cite{Weinzierl:2009ms}, and we additionally show that it is not necessary to explicitly exclude unviable patterns.

Given $N$ momenta $\{p_i\}$, with $\sum_i p_i=0$,  construct a set $S$ of the $2^N$ patterns of $N$ signs $s_i \in \{-1,1\}$ with $i=1,...,N$. 
We first view the signs as functions of the axis $n$, with
\be
s_i = \frac{p_i\cdot n}{|p_i\cdot n|}
\ee
We can then write
\be
\sum_i|p_i\cdot n| = \sum_i s_i p_i\cdot n
\ee
Clearly, this expression is maximised when $n = \sum_i s_i p_i / |\sum_i s_i p_i|$, so that
\be
\label{eq:viable}
\max_n \sum_i|p_i\cdot n| = \max_n n\cdot \sum_i s_i p_i =  \max_{S_{viable}} \frac{\sum_i s_i p_i}{|\sum_i s_i p_i|}\cdot \sum_i s_i p_i  =  \max_{S_{viable}} |\sum_i s_i p_i|
\ee
This statement is the one given in \cite{Weinzierl:2009ms}. The use of $S_{viable}$ instead of $S$ in eq.~(\ref{eq:viable}) is due to the fact that, as ref.~\cite{Weinzierl:2009ms} notes, only some of the sign patterns are allowed, meaning that once an axis $n$ has been determined, it holds true only for these allowed patterns that
\be
s_i = \frac{p_i\cdot n}{|p_i\cdot n|} \qquad \forall i \qquad \mathrm{with}~n = \frac{\sum_j s_j p_j}{|\sum_j s_j p_j|} 
\ee
We call these patterns `viable'.  For other patterns, one or more indices $i$ may exist for which this relation is not satisfied. We call them `unviable'.

We proceed to show that it is not necessary to exclude the unviable patterns when doing the maximisation over the sign patterns in eq.~(\ref{eq:viable}), since the unviable patterns never return a value larger than the maximum given by the viable ones. This means that $S_{viable}$ in eq.~(\ref{eq:viable}) can safely be replaced by $S$, leading to the form of the algorithm written in eq.~(\ref{eq:thrustalg}).

In order to see this, we now view the axis $n$ as a function of the signs, and write $n = \sum_i s_i p_i / |\sum_i s_i p_i|$. Suppose that we have an unviable pattern. This means that
\be
\label{eq:unviable}
\exists i = k \quad\mathrm{such~that} \quad s_k = - \frac{p_k\cdot n}{|p_k\cdot n|} 
\ee
(while $s_k$ just needs to ``differ'', since it is a sign it can only be the opposite sign). 
We can then write
\begin{eqnarray}
    \sum_i|p_i\cdot n|  &=& \sum_{i\ne k} |p_i\cdot n| + |p_k\cdot n| 
                        = \sum_{i\ne k} s_i p_i\cdot n - s_k p_k\cdot n \nonumber \\
                        &=& \sum_i s_i p_i\cdot n - 2 s_k p_k\cdot n \nonumber \\
                        &=& \sum_i s_i p_i\cdot \frac{\sum_j s_j p_j}{|\sum_j s_j p_j|} -2 \left(- \frac{p_k\cdot n}{|p_k\cdot n|} \right)p_k\cdot n \nonumber \\
                        &=& |\sum_i s_i p_i| + 2 |p_k\cdot n|
\end{eqnarray}
Considering that more than one `bad' index $k$ can exist with the property in eq.~(\ref{eq:unviable}), 
for un unviable pattern it therefore holds, at odds with eq.~(\ref{eq:viable}) valid for viable patterns,
\be
|\sum_i s_i p_i| = \sum_i|p_i\cdot n| - 2 \sum_{\{bad~k\}}|p_k\cdot n|
\ee
Since the second term on the r.h.s is always negative for an unviable pattern (and equal to zero for a viable one), maximising the terms of the equation above and using eq.~(\ref{eq:viable}) we can write
\ba
\max_S |\sum_i s_i p_i| &=& \max_n \sum_i|p_i\cdot n| - 2 \min \sum_{\{bad~k\}}|p_k\cdot n| \nonumber\\
&=& \max_{S_{viable}} |\sum_i s_i p_i| - 2 \min \sum_{\{bad~k\}}|p_k\cdot n|
\ea
i.e. $\max_S |\sum_i s_i p_i| \le \max_{S_{viable}} |\sum_i s_i p_i|$. The reverse inequality, $\max_S |\sum_i s_i p_i| \ge \max_{S_{viable}}|\sum_i s_i p_i|$, holds because $S_{viable} \subseteq S$. We have therefore
\be
\max_S |\sum_i s_i p_i| = \max_{S_{viable}} |\sum_i s_i p_i|
\ee 
which proves our statement.

\section{Minimum thrust for $N=5$ particles in three dimensions}
\label{app:n5conf}
\begin{figure}[t]
\includegraphics[width=0.325\textwidth]{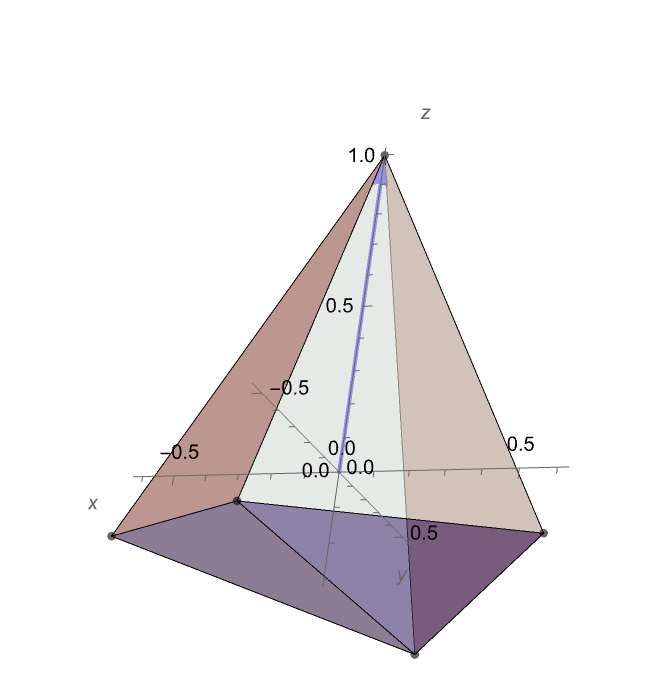}
\includegraphics[width=0.325\textwidth]{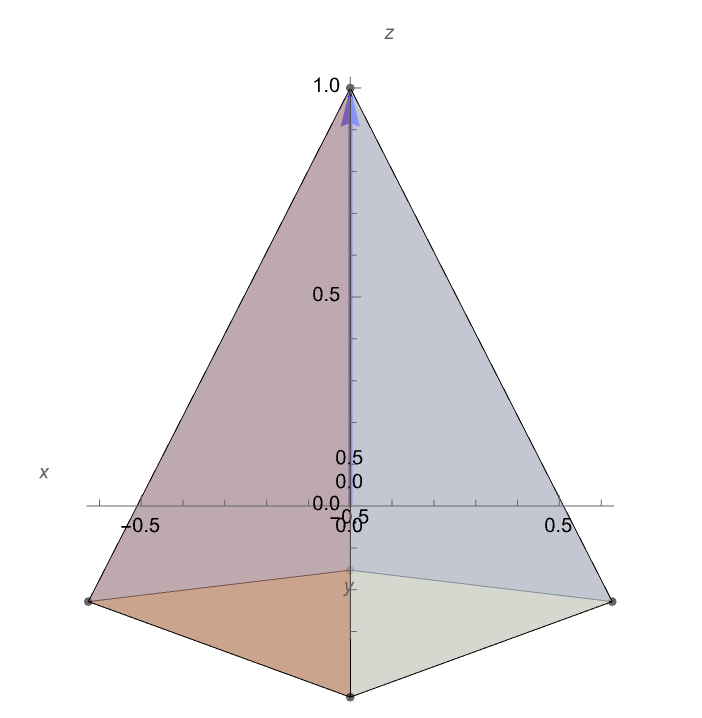}
\includegraphics[width=0.325\textwidth]{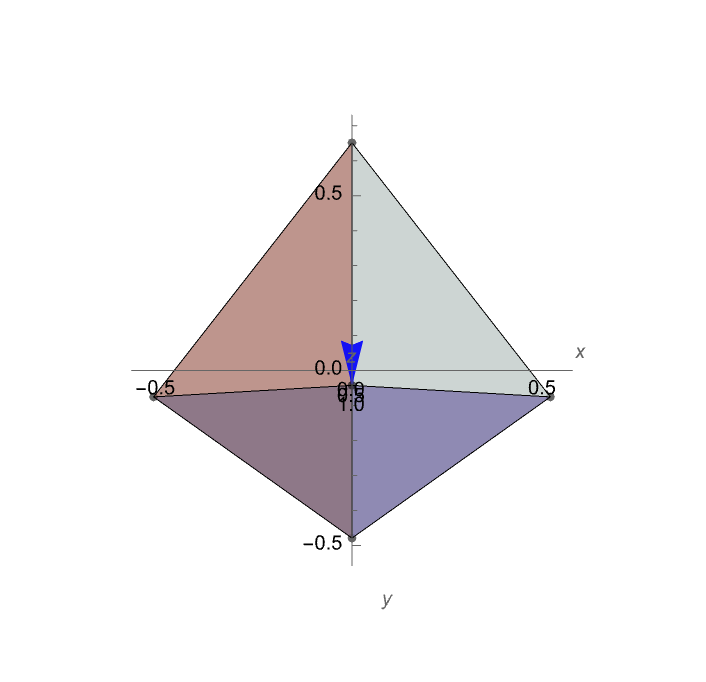}
\caption{\label{fig:n5conf} Configuration of minimum thrust for five particles. The second and third plot show a front and a top view respectively. The same configuration gives the same valua for the thrust with many other thrust axis directions.}
\end{figure}

The configuration with five particles corresponding to minimum thrust (and therefore to $\tau_{N=5}^{max}$), whose shape 
bears more than a passing resemblance to the Imperial Star Destroyer of Star Wars lore~\cite{starwars}, is found when determining the minimum thrust numerically and it is shown from different points of view in Figure~\ref{fig:n5conf}.

An exact expression for the coordinates of this configuration, and therefore for the value of minimum thrust itself, can be found using the algorithm given in Appendix~\ref{app:thrust}. 
After choosing arbitrarily the three dimensional coordinates of one particle, and accounting for momentum conservation, one is still left with $3(N-2)$ coordinates, i.e. nine for $N=5$ particles, which can be determined searching for the configuration that gives the minimum thrust according to the exact algorithm. 
Because the high dimensionality makes this procedure quite complex, 
hints taken from the configuration obtained via the numerical optimisation can be used to simplify it and guide the derivation.

One first observes that this configuration has a plane symmetry. Upon inspection one can also see that some of the coordinates are clearly either integer or half-integer numbers. This makes it possible to reduce the number of independent coordinates that need to be determined from nine to just two. At this point it becomes feasible to apply manually the algorithm in Appendix~\ref{app:thrust}, considering all sign patterns, and determining which one returns the minimum thrust. This question has an answer that depends on the values of the two remaining undetermined coordinates: the absolute minimum is generally at the intersection of two curves, corresponding to two different sign patterns. Solving for this intersection allows one to express one of the two missing coordinates as a function of the other, and the minimum thrust can then be determined using the standard minimisation technique of finding the zero of a derivative with respect to a single variable.

The value of the minimum thrust given by this procedure is
\be
T_{N=5}^{min} = \frac{2}{1 + \sqrt{1 + R^+}}
\ee
where $R^+ = 6.09140506144364...$ is the largest real root of the equation
\ba
&&78735300048 - 48783084768\, x + 20967607872\, x^2 - 9957160536\, x^3 \nonumber\\
&&+\, 3111202212\, x^4 - 554763700\, x^5 + 52381617\, x^6 - 2129402\, x^7 \nonumber\\ 
&&+\, 20333\, x^8 + 684\, x^9 + 4\, x^{10} = 0
\ea
The corresponding value of $\tau_{N=5}^{max}$ is
\be
\tau_{N=5}^{max} = 1- T_{N=5}^{min} = 0.45399486580028350678876698759830...
\ee
in perfect agreement with the one found numerically and given in Table~\ref{table:taumax}.

\section{Tests of EVT}
The effectiveness of Extreme Value Theory statistical analyses in predicting the true maximum can be tested in situations where such a value is known exactly, but it is not reached by the search procedure that is employed, the latter only returning a distribution of values smaller than the true maximum.

We have engineered two such situations:
\begin{enumerate}
\item
In three dimensions and for low-$N$ configurations, not employing at all a minimiser but simply generating random configurations. A situation equivalent to running a (seriously inefficient) minimiser multiple times can be created taking the minimum (or the maximum) of a large number of random configurations, and then doing the same with other batches of random configurations.
\item
In two dimensions and for odd values of $N$, where the theoretical values for $\tau_N^{max,2D}$ are easily known (see Appendix \ref{appA}), push $N$ to values large enough that the minimiser is unable to return multiple occurrences of the maximum value (within a reasonable time).
\end{enumerate}
In both cases, one can run our customary analyses (MLE and/or Bayesian) of the distributions that have been obtained, and check what values or intervals for the maximum are given, comparing them to the known values.

\begin{table}[t]
    \begin{center}
        \scriptsize
    \begin{tabular}{l l l l l l}
        \hline
        Configuration   
        & \begin{tabular}[c]{@{}c@{}}Exact\\maximum\end{tabular}  
        & \begin{tabular}[c]{@{}c@{}}Observed\\maximum\end{tabular}     
        & \begin{tabular}[c]{@{}c@{}}MLE\\endpoint\end{tabular}  
        & \begin{tabular}[c]{@{}c@{}}95\% credible\\ upper bound\end{tabular}            
        & \begin{tabular}[c]{@{}c@{}}99.7\% credible\\ upper bound\end{tabular}   \\
        \hline
        $N=3$ in 3D     & 0.3333333     & 0.333317          & 0.33333(4)    & 0.333401 & 0.333458\\
        $N=4$ in 3D     & 0.4226497     & 0.421023          & 0.42207(73)   & 0.422447 & 0.423254\\
        \hline
        $N=13$ in 2D    & 0.36182848    & 0.3618275         & 0.3618273(5)  & 0.3618290 & 0.3618305\\
        $N=15$ in 2D    & 0.36221518    & 0.3622134         & 0.3622132(7)  & 0.3622144 & 0.3622153\\

        \hline
    \end{tabular}
    \caption{\label{table:checks} Comparison between the known maxima $\tau_N^{max}$ in some specific configurations and the results returned by the statistical analyses. For each of the 3D cases 500 runs with 100000 random configurations each have been executed. For the 2D cases the Differential Evolution minimiser has been used, executing 400 and 500 runs for $N=13$ and $N=15$ respectively.}
\end{center}
\end{table}

Table~\ref{table:checks} shows the results of such comparisons. In the cases that we have considered (3 and 4 particles in three dimensions, 13 and 15 particles in two dimensions) the true, known maximum is not directly observed, but it is correctly included either in the 95\% upper credible bound or in the 99.7\% one.

\begin{table}[tp]
    \small
\begin{center}
    \begin{tabular}{l 
                    l 
                    c 
                    c}
        \hline
        $N$ & Numerical $\tau_N^{max,2D}$ 
            & Analytical $\tau_N^{max,2D} = 1- T_N^{min,2D}$           
            & Configuration \\
        \hline
        3   & 0.3333333333333 
            & $1- 2/3$ 
            & \raisebox{-1cm}{\includegraphics[width=2.5cm]{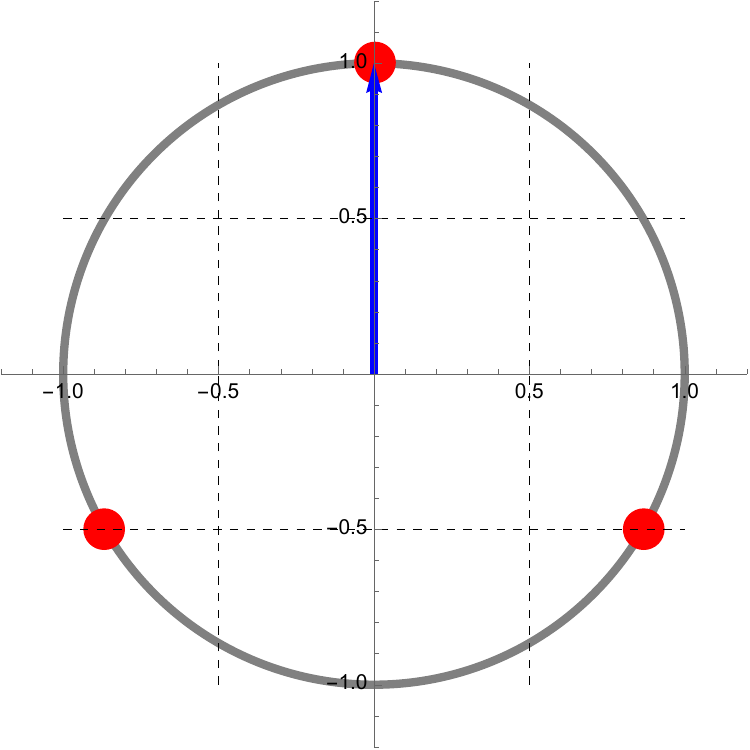}}\vspace{1mm} \\
        4   & 0.3410813774021 
            & $1- \frac{1}{2}(1-\sqrt{2} + \sqrt{3})$ 
            & \raisebox{-1cm}{\includegraphics[width=2.5cm]{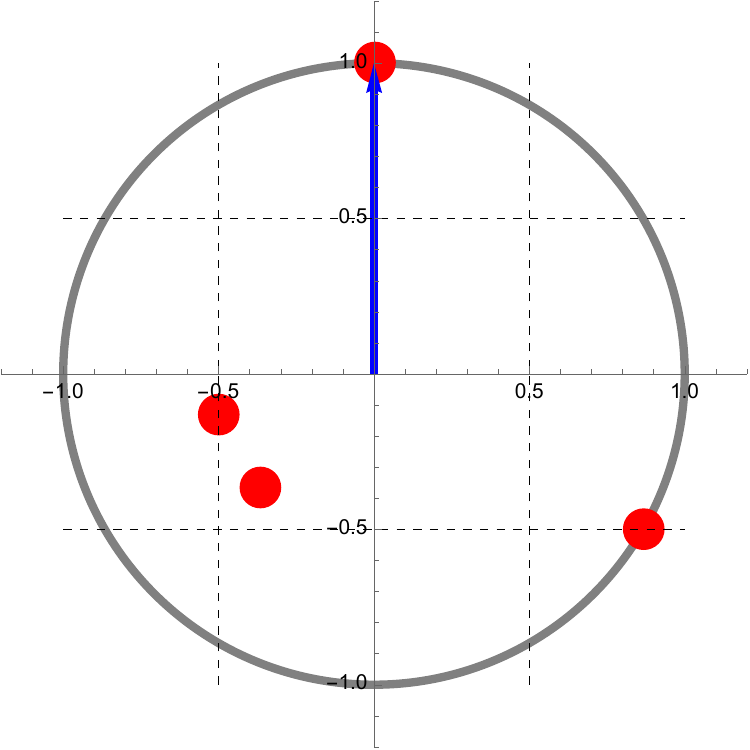}}\vspace{1mm}\\
        5   & 0.3527864045000 
            & $1- \frac{1+\sqrt{5}}{5}$ 
            & \raisebox{-1cm}{\includegraphics[width=2.5cm]{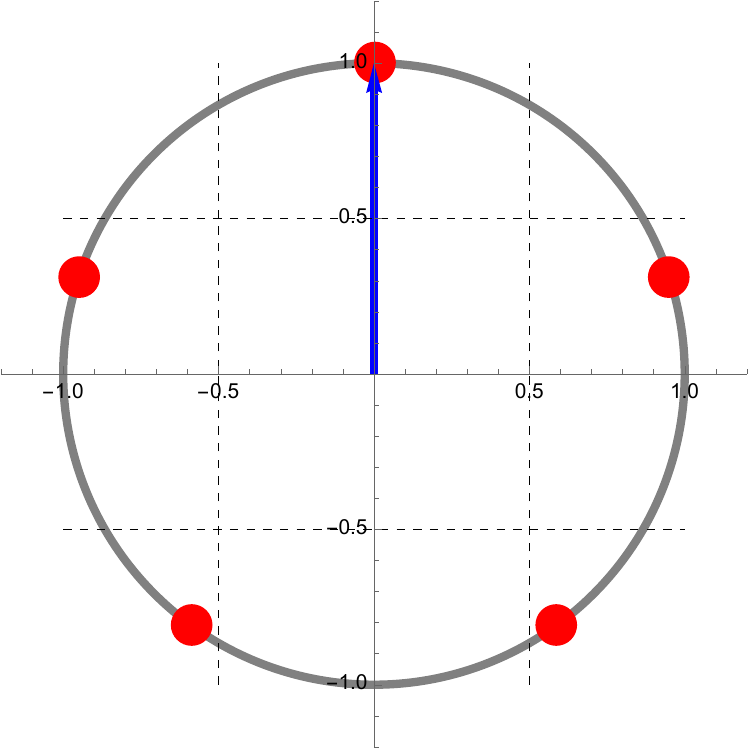}}\vspace{1mm} \\
        6   & 0.3560494491406   
            & $1- \frac{\sqrt{2}+\sqrt{6}}{6}$ 
            & \raisebox{-1cm}{\includegraphics[width=2.5cm]{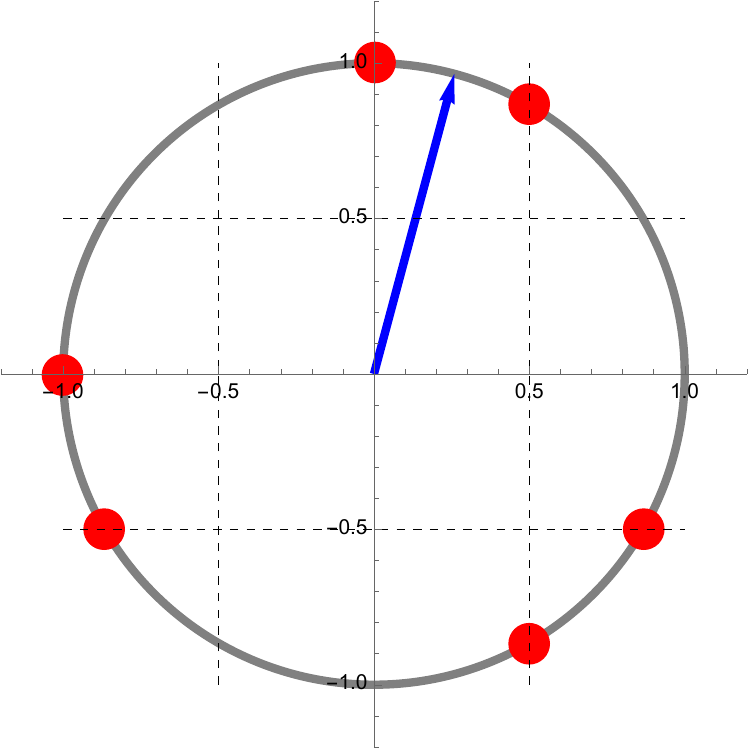}}\vspace{1mm} \\
        7   & 0.3580058275093 
            & $ 1- \frac{4}{21} (1 + \sqrt{7} \cos(\frac{1}{3}\arctan(3 \sqrt{3})))$   
            & \raisebox{-1cm}{\includegraphics[width=2.5cm]{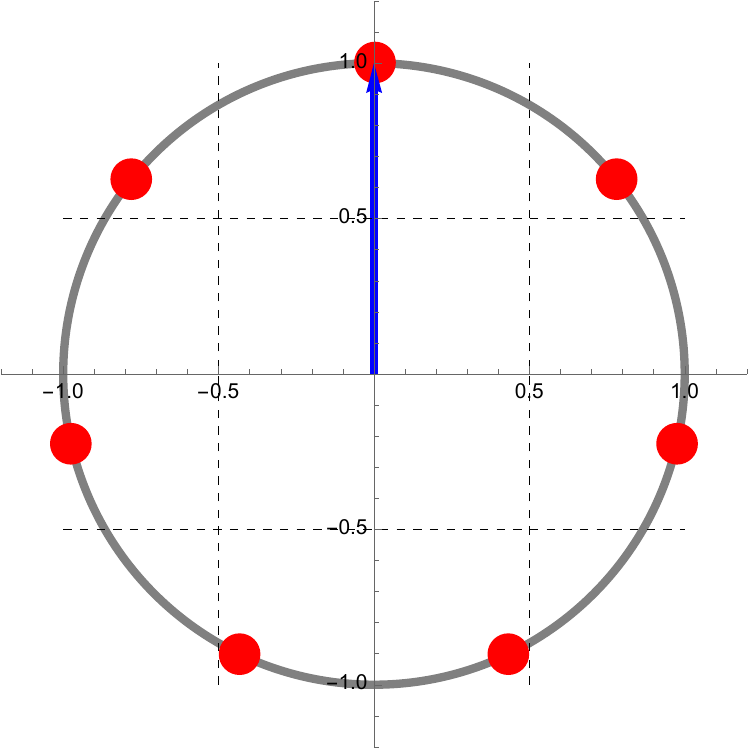}}\vspace{1mm} \\
        8   & 0.3592099590004 
            & -- 
            & \raisebox{-1cm}{\includegraphics[width=2.5cm]{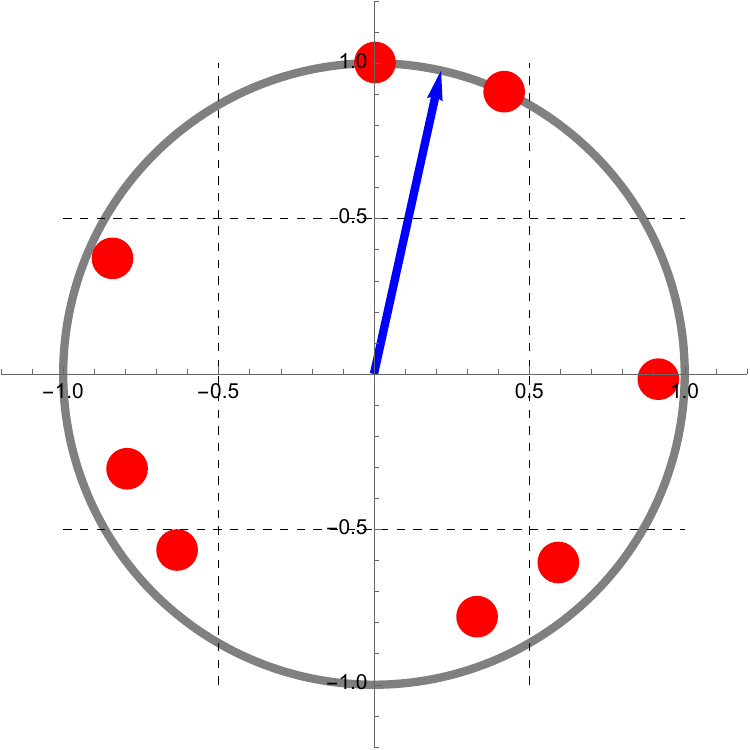}}\vspace{1mm} \\
        9   & 0.3601366129840 
            & $1-\frac{1}{9} (2 + 4\cos(\pi/9))$ 
            & \raisebox{-1cm}{\includegraphics[width=2.5cm]{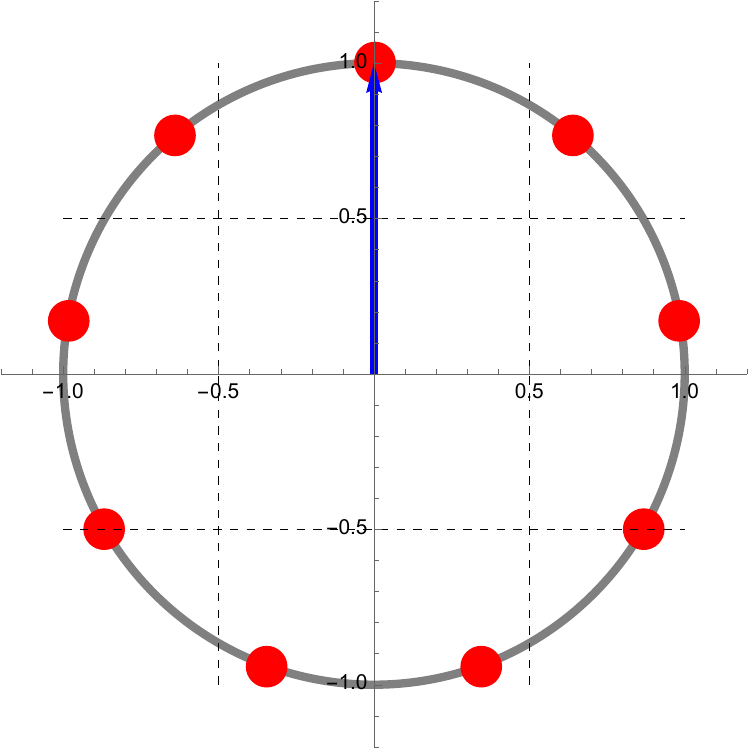}}\vspace{1mm} \\
        10  & 0.36075467785 
            & $ 1- \frac{1}{20} (\sqrt{2} (3 + \sqrt{5}) + 2 \sqrt{5 + \sqrt{5}})$ 
            & \raisebox{-1cm}{\includegraphics[width=2.5cm]{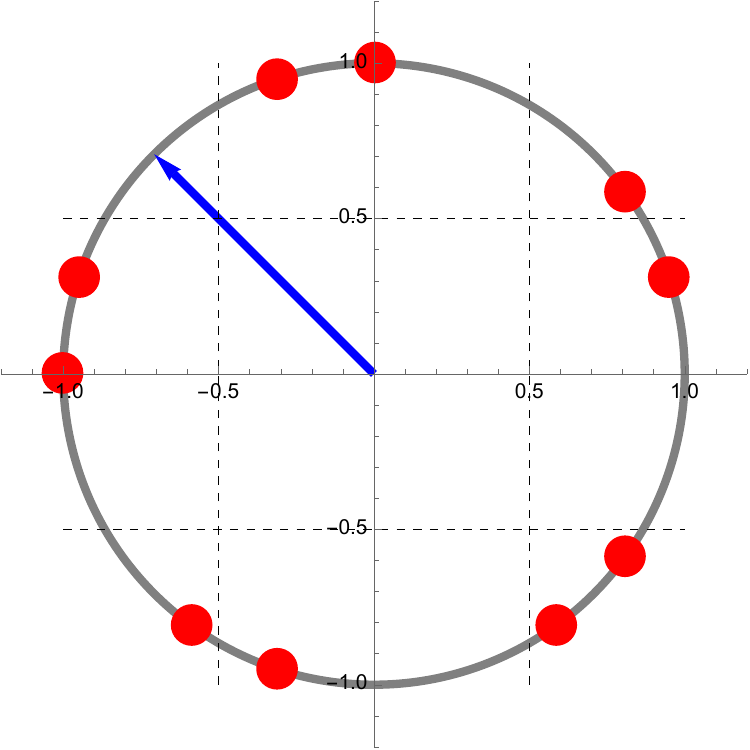}} \\
        \hline
    \end{tabular}
    \caption{\label{table:taumax2D} $\tau_N^{max,2D}$ values for planar configurations, as results of a numerical optimisation using Differential Evolution, and as analytical expressions where available, with the associated particle configurations. Note that multiple axis directions can usually give the same result for $\tau_N^{max,2D}$. 
    }
\end{center}
\end{table}

\section{Thrust in two dimensions}
\label{appA}

We show in Table~\ref{table:taumax2D} results obtained in two dimensions, with all particles in a plane. In this case the minimum theoretical value for thrust, with an infinite number of particles distributed uniformly on a circle, is $T_\infty^{min,2D} = 2/\pi$, leading to $\tau_\infty^{max,2D} = 1-T_\infty^{min,2D} \simeq 0.3633802276324187$.

Analytical results were obtained either by regularity considerations, then confirmed by the numerical analysis, or by reverse-engineering of the numerical results. It is worth noting that while configurations giving $\tau_N^{max}$ with an odd number of particles display the regularity that one may naively expect, this is not the case for configurations with an even number of particles. In particular, configurations with $N=4$ and with $N=8$ seem to have a maximal configuration where some of the particles are not on the unit-circle, at odds with all other cases.

Further inspection of the configurations leading to maximum $\tau$ (and minimum thrust) shows that some of them are built from combinations of lower-$N$ configurations. For instance, the maximal $N=6$ configuration is given by the superposition of two maximal $N=3$ configurations, with one of them rotated by an angle $\pi/6$ with respect to the other. The maximal $N=10$ configuration is also given by the superposition of two maximal $N=5$ configurations, with a relative rotation of an angle $\pi/10$.

Finally, while not shown in the table, for $N=12$ two equivalent maximal configurations exist, superpositions of two maximal $N=6$ configurations, one of which is rotated either by $\pi/12$ or by $\pi/4$. They both give
\be
\tau_{N=12}^{max,2D} = 1-\frac{1}{6} \sqrt{4+3 \sqrt{\frac{3}{2}}+\frac{5}{\sqrt{2}}+2 \sqrt{3}} = 0.3615585353716343...
\ee

\section{Thrust in $d$ dimensions}
In line with the beloved tradition of certain fields of physics of working in dimensions that may have little to do with reality, we give here some results for a number of spatial dimensions equal to four or higher.

\underline{\sl Assuming} that, when the number of particles is equal to the number of dimensions plus one, the configuration that gives the minimum thrust in $d$ dimensions is the regular $d$-simplex (also known as a regular $d$+1-cell) inscribed in a unit $d$-sphere, as is the case in two dimensions (three particles, an equilateral triangle inscribed in a unit-circle) and in three dimensions (four particles, a regular tetrahedron inscribed in a unit-sphere), we give the result for  $d+1$ particles in $d$ dimensions.
One finds
\be
\tau_{N=d+1}^{max,dD} = 1- T_{N=d+1}^{min,dD} = \left\{\begin{tabular}{l l}
    $1-\frac{\sqrt{d+2}}{d+1}$  &\qquad for $d$ even \\[10pt]
    $1-\frac{1}{\sqrt{d}}$      &\qquad for $d$ odd 
\end{tabular} \right.
\ee

For $d=2$ and $d=3$ one recovers the known results for three and four particles in 2- and 3-dimensional space respectively.


\begin{thebibliography}{9}

\bibitem{Brandt:1964sa}
S.~Brandt, C.~Peyrou, R.~Sosnowski and A.~Wroblewski,
Phys. Lett. \textbf{12} (1964), 57-61
doi:10.1016/0031-9163(64)91176-X

\bibitem{Farhi:1977sg}
E.~Farhi,
Phys. Rev. Lett. \textbf{39} (1977), 1587-1588
doi:10.1103/PhysRevLett.39.1587

\bibitem{Aglietti:2025jdj}
U.~G.~Aglietti, G.~Ferrera, W.~L.~Ju and J.~Miao,
Phys. Rev. Lett. \textbf{134} (2025) no.25, 251904
doi:10.1103/dv7n-qvyp
[arXiv:2502.01570 [hep-ph]].

\bibitem{Monni:2011gb}
P.~F.~Monni, T.~Gehrmann and G.~Luisoni,
JHEP \textbf{08} (2011), 010
doi:10.1007/JHEP08(2011)010
[arXiv:1105.4560 [hep-ph]].



\bibitem{Weinzierl:2009ms}
S.~Weinzierl,
JHEP \textbf{06} (2009), 041
doi:10.1088/1126-6708/2009/06/041
[arXiv:0904.1077 [hep-ph]].

\bibitem{Banfi:2000si}
A.~Banfi, G.~Marchesini, Y.~L.~Dokshitzer and G.~Zanderighi,
JHEP \textbf{07} (2000), 002
doi:10.1088/1126-6708/2000/07/002
[arXiv:hep-ph/0004027 [hep-ph]].

\bibitem{numpy}
https://numpy.org

\bibitem{de}
R.~Storn, and K.~Price, Differential Evolution - a Simple and Efficient Heuristic for Global Optimisation over Continuous Spaces, Journal of Global Optimization, 1997, 11, 341 - 359.

\bibitem{scipy}
Pauli Virtanen {\it et al.},
Fundamental Algorithms for Scientific Computing in Python. Nature Methods, 17(3), 261-272. DOI: 10.1038/s41592-019-0686-2. https://scipy.org

\bibitem{cmaes}
Nikolaus Hansen and Andreas Ostermeier,
Completely derandomized self-adaptation in evolution strategies. 
Evol. Comput., 9(2):159–195, June 2001. URL: http://dx.doi.org/10.1162/106365601750190398, doi:10.1162/106365601750190398.

\bibitem{pymoo} 
https://pymoo.org/

\bibitem{pypi}
https://pypi.org/



\bibitem{pymc}
https://www.pymc.io

\bibitem{starwars}
https://www.starwars.com/databank/imperial-star-destroyer


\end{thebibliography}
\end{document}